\documentstyle[12pt]{article}
\def\singlespace {\smallskipamount=3.75pt plus1pt minus1pt
                  \medskipamount=7.5pt plus2pt minus2pt
                  \bigskipamount=15pt plus4pt minus4pt
                  \normalbaselineskip=15pt plus0pt minus0pt
                  \normallineskip=1pt
                  \normallineskiplimit=0pt
                  \jot=3.75pt
                  {\def\smallskip {\vskip\smallskipamount}}
                  {\def\medskip   {\vskip\medskipamount}}
                  {\def\bigskip   {\vskip\bigskipamount}}
                  {\setbox\strutbox=\hbox{\vrule 
                    height10.5pt depth4.5pt width 0pt}}
                  \parskip 7.5pt
                  \normalbaselines}
\def\middlespace {\smallskipamount=5.625pt plus1.5pt minus1.5pt
                  \medskipamount=11.25pt plus3pt minus3pt
                  \bigskipamount=22.5pt plus6pt minus6pt
                  \normalbaselineskip=22.5pt plus0pt minus0pt
                  \normallineskip=1pt
                  \normallineskiplimit=0pt
                  \jot=5.625pt
                  {\def\smallskip {\vskip\smallskipamount}}
                  {\def\medskip   {\vskip\medskipamount}}
                  {\def\bigskip   {\vskip\bigskipamount}}
                  {\setbox\strutbox=\hbox{\vrule 
                    height15.75pt depth6.75pt width 0pt}}
                  \parskip 11.25pt
                  \normalbaselines}
\def\doublespace {\smallskipamount=7.5pt plus2pt minus2pt
                  \medskipamount=15pt plus4pt minus4pt
                  \bigskipamount=30pt plus8pt minus8pt
                  \normalbaselineskip=30pt plus0pt minus0pt
                  \normallineskip=2pt
                  \normallineskiplimit=0pt
                  \jot=7.5pt
                  {\def\smallskip {\vskip\smallskipamount}}
                  {\def\medskip   {\vskip\medskipamount}}
                  {\def\bigskip   {\vskip\bigskipamount}}
                  {\setbox\strutbox=\hbox{\vrule 
                    height21.0pt depth9.0pt width 0pt}}
                  \parskip 15.0pt
                  \normalbaselines}

\include{dspace12}
\def\be{\begin{equation}}
\def\ee{\end{equation}}
\def\bea{\begin{eqnarray}}
\def\eea{\end{eqnarray}}

\def\sect #1{\setcounter{equation}{0}}

\begin{document}
\begin{center}
{\Large {FIELD OF A RADIATION DISTRIBUTION}}
\end{center}
\vspace{1.0in}
\vspace{12pt}

\begin{center}
{\large{ I. H. Dwivedi\footnote{E-mail : dwivedi@tifrvax.tifr.res.in}\\
Theoretical Astrophysics Group\\
Tata Institute of Fundamental Research\\
Homi Bhabha Road, Colaba, Bombay 400005, India.}}
\end{center}

\vspace{1.3in}

\begin{abstract}

General relativistic spherically symmetric matter field with a vanishing 
stress energy scalar is analyzed. Procedure for generating exact solutions 
of the field equations for such matter distributions is given. It is 
further pointed out that all such type I spherically symmetric fields 
with distinct eignvalues in the radial two space can be treated as a 
mixture of isotropic and directed radiations. Various classes of exact 
solutions are given. Junction conditions for such a matter field to
the possible exterior solutions are also discussed.

\end{abstract}

\noindent
{\bf I. INTRODUCTION}

Physical aspects of nature can be formulated in the form of a mathematical 
model comprising of a set of differential equations. Solutions to this 
set of differential equations are used not only to confirm the validity 
of the proposed theory by  way of comparing its prediction for a particular 
known phenomena in nature, but also to predict possibly new unknown or even 
yet to be discovered natural phenomena. General relativity, despite being 
the most elegant mathematical theory of gravitation, has one of the
most complicated set of differential equations and getting exact solutions 
describing a realistic physical scenario is a difficult task. 
Though after the discovery of the first exact solution, namely the 
Schwarzschild solution, numerous sets and classes of exact solutions of 
the field equations are given in the literature$^1$, still only very few
of them are physically realistic. Perhaps this may be the reason that 
despite being accepted as the best theory of gravitation and the fact
that gravity is the most dominant force at present stage in our 
universe the tests of relativity theory as a true theory of gravitation 
are basically three coming from the first discovered and most widely
used vacuum Schwarzschild solution. The study of various solutions
of the fields equations from the mathematical point of view has its own merit
and from that prospective there is a sea of solutions to the field equation.
However, we still do not have exact solutions for most physical situations.
The very few exact solutions that do describe some physically realistic
situations are  Schwarzschild space-time and Ker space-time
in vacuum, Friedmann models in cosmology, Tolman-Bondi dust solutions and
the Vaidya space-time in the case of non vacuum solutions.

In order to solve field equations basically one can use either g-method 
or T-method. In g-method  metric functions $g_{ab}$ are assigned and 
then the energy momentum tensor $T_{ab}$ is obtained by mere differentiation
via field equations. Thus in this regard every metric is an 
exact solution of the field equations. The problem with this method
is that resulting $T_{ab}$ may and very likely do not describe a physically
realistic distribution of matter. The T-method is most promising in the
sense that one starts with a physically realistic situation by
prescribing a suitable $T_{ab}$ and the field equations are then integrated
to obtain the geometry of the spacetime via $g_{ab}$. However due to the
high degree of non linearity of the differential equations involved, 
in very few cases exact solutions of the field equations has been obtained 
using this method. In this paper we wish to analyze the field equations
from this perspective by prescribing stress energy tensor and 
use the T-method to obtain solutions of the field equations.

The vacuum Schwarzschild solution gives the gravitational field outside a 
spherically symmetric distribution of matter. However 
most of the stars radiate energy in the form of lightlike particles
and thus the immediate space outside the star is filled with electromagnetic
radiations (photon ). The Vaidya spacetime$^2$ is used to
describe the field of a radiating star. The stress energy tensor in Vaidya
solution is of the form
$$T^{ab}=\sigma k^ak^b$$
which describes only the radially directed radiations. 
$\sigma$ is regarded as the energy flux density of the directed radiations 
in some frame depending on the normalization the vector $k^a$ which
is tangent to null geodesics. In spherically symmetric spacetime it
can be regarded as outgoing radiations with momentum only in
radial direction and with $\sigma =p$, $p$ being the radial pressure. Vaidya
spacetime had been quite extensively used not only to study stellar 
objects but also in connection with spherical collapse
and formation of naked singularities$^3$. 
However, in nature the space outside radiating stars is not filled 
only by the directed radiations but in fact there is a presence of
isotropic radiations surrounding the immediate space out side the star. Such 
a gas of photons (isotropic radiations) may originate either from the 
star itself or from some outside sources. The energy density of such a 
photon gas (isotropic radiations) may be small or may even
be decreasing much faster than the flux of directed radiations, 
which reach the distant observers.

The energy momentum tensor of the Vaidya spacetime is actually part of a 
larger class of energy momentum tensors which have a vanishing
stress energy scalar namely $T=T^a_a=0$. In fact type I matter
fields$^4$ , which account for all the known and observed physically 
realistic fields (except directed radiations which is type II)
, with a vanishing $T$ can be expressed as mixture of both isotropic
and directed radiations as we point out in this paper.
The aim of this paper is to analyze the field equations for spherically
symmetric matter distributions with vanishing stress energy scalar
and to find solutions in such cases where spacetime
is filled with isotropic as well as directed radiations.

Thus the solutions  obtained in this paper could be significant in
the context of a gravitational collapse scenario. It is believed that in 
the late stages of collapse matter may disintegrate due to extreme 
temperature, densities and pressures into photons or lightlike particles 
in the form of a gas of lightlike particles. This may consist of both or 
either isotropic and anisotropic radiations. It could be argued that this 
photon gas may then disperse and thus no singularity may form as a result 
of collapse. We do, however, find in this paper that certain
exact solutions describing a photon cloud as 
considered here may still develop a singularity in the spacetime. 
As expected all the static photon gas
solutions with only isotropic radiations do not match with the 
Schwarzschild vacuum solution but interestingly some solutions
representing a mixture of a isotropic photon gas and directed radiations 
as considered in this paper can be matched to the vacuum 
exterior only at the sphere of radius $14 M_s/4$ where $M_s$ is the 
Schwarzschild Mass of the photon cloud.

The paper is organized as follows. In section II we consider the basic field
equations for our model of spacetime describing the spherically symmetric
matter distribution with vanishing $T$. A procedure for solving and
obtaining solutions of the field equations with detailed discussion
on type I fields is given in this section. In Section III we give
a method for generating exact solutions of the field equations in this
scenario. We also explore some explicit exact solutions describing various
density distributions of the photon cloud in this section. Section IV
is devoted to the matching conditions of solutions thus obtained to
possible exterior metric out side the radiation zone. We discuss in
appendix A the stress energy tensor with vanishing $T$ and point out that
the same can be expressed as a mixture of isotropic radiations with
diffusive terms as well as directed radiations. Appendix B very briefly
gives the geometric quantities associated with the metric written
in null coordinates.
 
\noindent
{\bf II. Field Equations and Radiation distributions}

Stress energy tensor of a spherically symmetric spacetime with
a vanishing stress energy scalar $T=T^a_a=0$ (Ricci curvature scalar $R=0$)
can always be expressed as (see appendix A)
\begin{equation} T^{ab}=T^{ab}_{(\hbox{pht gas})}+ T^{ab}_{({\hbox
{dirct rad}})}+T^{ab}_{(diffusi)}\end{equation}
where
\begin{equation}T^{ab}_{\hbox{(pht gas)}}=(\rho+p)v^av^b+pg_{ab},
\quad p={\rho\over 3}
\end{equation}
\begin{equation}T^{ab}_{\hbox{(dirct rad)}}=\sigma k^ak^b\end{equation}
\begin{equation}T^{ab}_{(diffusi)}=q^av^b+q^bv^a\end{equation}
Here $T^{ab}_{(\hbox{pht gas})}$ represents the stress energy tensor 
of the isotropic radiations (photon gas)$^5$, $\rho$ is the energy density
of the isotropic radiations as measured in the rest frame of the 
timelike vector $v^a$ (i.e. $v^av_a=-1$). 
$T^{ab}_{\hbox{(dirct rad)}}$ is the energy tensor of
the directed radiations with vector
$k^a$ tangent to null geodesics and $\sigma $ is the energy density
of the directed radiations. $q^a$ is a spacelike vector normal
to the timelike vector $v^a$ and is treated as the heat flux vector and 
$T^{ab}_{(diffusi)}$ represent contribution of diffusive terms 
in the stress energy tensor.  

Using the spherical symmetry and adopting the coordinates 
such that $k^a\propto \delta^a_1$ (i.e. null coordinates) the 
metric describing the space-time is then given by
\begin{equation}ds^2=-Adu^2\pm 2e^{2\beta}dudr +r^2(d\theta^2+
\sin^2{\theta}d\phi^2)
,\quad u,r,\theta,\phi=0,1,2,3\end{equation}
Vector $k_a=\delta _a^0$ is the propagation vector of the directed radial
radiations, and $A=A(u,r)$ and $\beta=\beta(u,r)$ are functions of 
$u$ and $r$, $\pm$ relate to the use of either retarded of advanced time
and is used to distinguish ingoing or outgoing  
radiations.

Stress energy tensor has only $T_{00}$ , $T_{01}$ , $T_{11}$ 
and $T_{33}=sin^2{\theta}T_{22}$ as non 
vanishing components. Since $ T^a_a=0$ field 
equations for the space-time described by the metric (5) basically
reduce to the following four equations
\begin{equation}R_{00}={\it k}T_{00},\quad R_{11}={\it k}T_{11},
\quad R_{22}={\it k}
T_{22},\quad R=0\end{equation}
where ${\it k}$ is the gravitational constant and $R$ is the 
Ricci scalar curvature. Once the above equations are
satisfied equation $R_{01}={\it k}T_{01}$ is automatically satisfied.
Using the metric (5), field equations (6) become (see appendix B
for various geometric quantities and expressions for Ricci tensor)
\begin{equation}{\it k}T_{11}={4\beta '\over r}\end{equation}
\begin{equation}{\it k}T_{22}= 1-e^{-2\beta}(rAe^{-2\beta})'\end{equation}
\begin{equation}{\it k}T_{00} 
=\mp e^{2\beta}\dot{({Ae^{-4\beta}\over r})}\pm Ae^{-2\beta}\dot\beta'
+{Ae^{-4\beta}\over 2}(A''+A'({4\over r}-2\beta')
)\end{equation}
\begin{equation}R=0\Rightarrow r^2A''+2rA'(2-r\beta')+2A(
1-4r\beta')=
2e^{4\beta}\mp r^2e^{
2\beta}4\dot\beta'\end{equation}
where $(\dot{ })$ and $(')$ represent partial derivatives with respect to
$u$ and $r$ respectively. The procedure for solving the above field
equations is straight forward as follows. Since there are only four 
equations to be satisfied and there are five unknowns namely 
$A, \beta, \rho, \sigma$ and $q^2=q^aq_a$ there is a degree of freedom 
in the choice of one function. From equation (1) and (7) follows that 
$\beta'$ is related to the density of isotropic radiations $\rho$ and 
the $q$. Therefore in general one chooses $\beta=\beta(u,r)$ in a way 
desired for suitable density distribution of the photon gas or otherwise 
as the case may be. Equation (10) which is a second order ordinary 
differential for function $A$ in variable $r$ is then solved 
to find $A=A(u,r)$. The rest of the remaining three equations (7) to (9)
then determine the three unknowns $T_{00},T_{11}$ and $T_{22}$, which give
the energy density of the isotropic radiations $\rho$, 
energy density of the directed radiations $\sigma$ and the heat flux $q$. 

We wish to discuss mainly in this paper type I spherically symmetric matter 
fields $^4$ in the context of exact solutions, since all observed fields 
are of this type except for type II which represent directed radiations 
only. Any type I matter field ( i.e. $T^{ab}$) with vanishing scalar $T$ 
can be cast as (see equation (70) in appendix A) 
\begin{equation} T^{ab}={4\rho\over 3}v^av^b +{\rho\over 3}g^{ab} +
\sigma k^ak^b\end{equation}
except the special cases where the eignvalues
corresponding to the eignvectors in the radial two space are equal in 
magnitude(see equations (71) and (72) in appendix A). We would mention
here for the sake of completeness that only other
physically reasonable matter fields are of type II and  the stress energy
tensor for this case is either Vaidya null dust given in equation
(73) or photon gas with diffusion terms as given in equation (74). All other
types of matter fields are not considered physically reasonable as they
necessarily violate energy conditions.

We would focus our discussion on type I unless specified otherwise 
and we would use stress energy tensor given in (11) when giving 
explicit exact solutions. The stress energy tensor $T^{ab}$ in (11) 
represents type I matter fields and the weak energy conditions are satisfied 
provided
\begin{equation}\rho \ge 0,\quad \sigma \ge 0\end{equation}
Let us now consider the field equation for a cloud of gas
consisting of both isotropic and directed radiations as
described by the stress energy tensor given in (11). 
For space-time given by (5) timelike
vector $v_a=(v_0,v_1,0,0)$ ($v_1\ne 0$) satisfies
\begin{equation}Ae^{-4\beta}(v_1)^2\pm 2e^{-2\beta}v_0v_1=-1\end{equation}
Note that for $T^{ab}$ as in (11)
function $\beta'$
is related to the density distribution of isotropic radiations in the 
space-time, for example if $\beta'=0$ then from equation (7) and
(11) it follows that $T_{11}=0\rightarrow \rho =0$ and the cloud does 
not have any isotropic photon gas and is all directed radiations and 
the solution is given by  Vaidya,s radiating star$^2$. 
Hence one chooses $\beta'$ as required for the suitable density distribution 
of the isotropic radiations and integrates equation (10) to find $A$.
The energy density of the isotropic photon gas
as measured in the rest frame of the timelike vector $v^a$ 
and of the directed radiations 
in the photon cloud are then given by  
equations (7) to (9) in terms of $\beta$ and $A$. 
We further have for $\rho$, $\sigma$, and $v_1$ using equations (5), (11),
(13) and (7) to (10) in terms of $A$ and $\beta$
\begin{equation}{{\it k}\rho r^2\over 3}={\it k}T_{22}= 
1-e^{-2\beta}(rAe^{-2\beta})'\end{equation}
\begin{equation}{\it k}\rho(v_1)^2={3\over 4}{\it k}T_{11}=
{3\beta '\over r}\end{equation}
\begin{equation}\sigma=T_{00}-
{\rho\over 3}(4v_0^2-A)
=\pm e^{2\beta}{\dot {(rAe^{-4\beta})}\over {\it k}r^2}-
{\rho
\over 3g^{11}(v_1)^2}
(1-(g^{11}(v_1)^2)^2)\end{equation}

{\bf IV. Exact Solutions}

As mentioned in the preceding section, one can solve the field equations 
by specifying the function $\beta(u,r)$ 
and integrating the ordinary second order linear differential equation (
10) to get $A(u,r)$. Rest of the quantities are then calculated by
differentiating $\beta$ and $A$ from equations (7 ) to (9) and (13).

The simple procedure to find an exact solution is as follows. 
Consider the homogeneous part of the
differential equation  (10) for $A=A(u,r)$ which is given by
\begin{equation}y''+{y'\over r}(4-2r\beta')+{2y\over r^2}(1 -4r\beta')
=0\end{equation}
This is an ordinary second order linear differential equation and variable
$u$ and functions of $u$ which appear in $\beta '(u,r)$ are treated as 
constant in the differential 
equation. Let $y=y_1(r)$ be a solution of the differential equation (17). The
general solution $A=A(u,r)$ of differential equation (10) is then given by
\begin{equation}A(u,r)=M_1(u)y_1+M_2(u) y_2+A_p\end{equation}
\begin{equation}y_2=y_1\int{{e^{2\beta}\over r^4 y_1^2}dr}\end{equation}
\begin{equation}A_p(u,r)
=y_1\int{\left(\int{(e^{2\beta}r^2y_1\mp 
r^4\dot\beta'y_1dr}\right ){e^{2\beta
}dr\over (r^2y_1)^2}}\end{equation}
Where $M_1(u)$ and $M_2(u)$ are arbitrary functions of $u$ ,$y_2$ is 
another solution of the homogeneous equation and $A_p$ 
represents the particular solution of equation (10). Thus $A(u,r)$ in above
equation gives the general solution of the field equations for a given
$\beta=\beta (u,r)$. In case $\beta '=0$ the solution above reduce
to charged null dust solution$^6$ (i.e. charged Vaidya solution)

An interesting way to generate exact solutions 
for different density distributions is by selecting the function $y_1=
y_1(u,r)$ and solving for $\beta$ in equation (17) that is
\begin{equation}r\beta '={y_1''+{4y_1'\over r}+
{2y_1\over r^2}\over {2y_1'\over r}+{8y_1
\over r^2}}\end{equation}
\begin{equation}\Rightarrow e^{2\beta}={ry'+4y\over r}
\exp(\int{{3dr\over r(ry'+4)}})
\end{equation}  
Thus for any given $y_1(u,r)$ $\beta$ is known from the above equation (21)
and then $A$ is given
by the equation (18). Rest of the
unknowns namely $T_{00},T_{11}, T_{22}$ are immediate from
equations (7) to (9) for the general $T^{ab}$ in (1),
and in case of type I matter field   
$\rho(u,r),v_1$, and $\sigma(u,r)$ are given by (14) to (16).

We now consider few examples of exact solutions of physical interest for
a photon cloud given by the energy tensor in (11)
and to clarify the procedure.

{\bf 1. $ y_1=r^n, \quad n\ne -4 $}

Let us put $y_1=r^n$, we get $\beta$ from (22)
\begin{equation}e^{\beta(u,r)}=({r\over r_o})^m,
\quad m={(n+1)(n+2)\over 2(n+4)}\end{equation}
Equations (19) and (20) give 
\begin{equation}y_2=r^l,\quad l=-{4n+10\over n+4}\end{equation}
\begin{equation}A_p={2e^{4\beta}\over (2m+n+3)(4m-n)}
\mp {e^{2\beta}(n^2+8n+10)\dot n
\over 2(n+4)^2}\end{equation}.
The general solution for $A$ is then
\begin{equation}A(u,r)=M_1(u)y_1+M_2y_2+{2e^{4\beta}\over a_o}
\mp b_oe^{2\beta}\end{equation}.
\begin{equation} a_o=(2m+n+3)(4m-n),\quad
b_o={(n^2+8n+10)\dot n
\over 2(n+4)^2}\end{equation}

The energy density of the isotropic and directed radiations are
$${{\it k}\rho r^2\over 3}=1-(M_1(u)(n-2m+1)y_1e^{-4\beta}
+M_2(u)(l-2m+1)y_2e^{-4\beta}$$
\begin{equation}+
{2(4m+1)\over a_o}
\mp e^{-2\beta}b_o)\end{equation}
$${\it k}\sigma=\pm {e^{2\beta}\over r}\left (y_1\dot 
{(M_1(u)e^{-4\beta})}+
y_2\dot 
{(M_2(u)e^{-4\beta})}
+\dot {({2\over a_o})}
\mp \dot {(b_oe^{-2\beta})}\right)$$
\begin{equation}-{{\it k}^2\rho^2\over 9mg^{11}}
(1-({3mg^{11}\over {\it k}\rho r^2})^2)\end{equation}
\begin{equation}(v_1)^2={3m \over {\it k}\rho r^2}
\end{equation}
Thus the solution involves four arbitrary functions of $u$ namely
$M_1(u)$, $M_2(u)$, $n(u)$ and $r_o(u)$. We next consider few solutions
for a given $\beta(u,r)$

{\bf 2}. $\beta=-\mu(u)/ r^n$

In this section we mainly wish to obtain exact solutions which are
either asymptotically flat in the sense that the metric becomes Minkowskian as
$r\rightarrow \infty$ or reduce to Vaidya spacetime for $r>>0$. 
Furthermore these spacetime would reduce to Vaidya
space-time in case $\mu =0$ as the isotropic energy density of photon gas
would vanish.  
We give below some exact solutions for different configurations. As
pointed out above, once a solution of the homogeneous equation (17) 
given as $y=y_1(r)$, the specification of 
all the unknowns namely $\rho,\sigma, A, v_1$ as pointed out earlier, 
becomes immediate from
equations (13) to (16) and (18). Hence we give below few examples of
exact solutions thus obtained.

\begin{equation}a) \beta=-{-\mu\over r}, \quad y_1={1\over r^2}
(1+{2\mu\over r}+{2\mu^2\over 3r^2})\end{equation}

\begin{equation}b) \beta=-{-\mu\over r^2}, \quad y_1={1\over r^2}
(1+{4\mu\over r^2})\end{equation}

\begin{equation}c) \beta=-{-\mu\over r^3}, \quad y_1={1\over r}
(1+{3\mu\over r^3})\end{equation}

For a general form of $\beta$ which is vanishing as $r\rightarrow \infty$
one can find the power series solution of the
homogenous equation (17) in the form of $y=y_1(r)$. 
The solution of the field equation is then immediate
by the determination of $A$ from equations (18) which is given in 
terms of $\beta (u,r)$ and
$y_1$.

{\bf 3. Static Photon Cloud}

For static solutions $\beta=\beta(r)$ and $A=A(r)$ hence in all previous
solutions reduce to static solutions by putting function depending on $u$
as constants. However in static case energy densities of both the directed
and isotropic radiations are related as we obtain from equation (16)
$${\it k}\sigma={mg^{11}\over \ r^4}-{({\it k}\rho)^2\over 9mg^{11}}$$
In cases where the photon cloud does not contain any isotropic
radiations and is static note that in such cases $\rho=0$ implies
from (16) $\sigma=0$ and the spacetime does not have any radiations.
We next consider exact solutions 
in cases where the radiation cloud does
not consists of any directed radiations$^8$ and is static i.e.
$\dot \beta =\dot A=0$. Requiring additionally that energy density 
of the directed radiations $\sigma=0$
we get from equation the above equation for that
\begin{equation}\sigma=0 => 1-g^{11}(v_1)^2=0\end{equation}
Hence from equations (10) and (13) to (16) we get
\begin{equation}e^{3\beta}=Y',\quad A={Ye^{\beta}\over r}\end{equation}
\begin{equation}Y''Y^3=c_or^4(Y')^{4\over 3}\end{equation}
where $c_o$ is a constant.  Thus any solution of equation (35) gives a 
solution representing a cloud of static photon sphere 
(pure isotropic radiations). 
For one such solution  note that
$Y=const. r^{{7\over 4}}$ satisfies equation (36) and hence we 
get the solution as
\begin{equation}e^{\beta}=({r\over r_o})^{{1\over 4}}\end{equation}
\begin{equation}A={4r\over 7r_o}\end{equation}
\begin{equation}\rho={3\over 7 {\it k}r^2},\quad \sigma=0,\quad v^a=
\sqrt{{7r_o\over 4r}}\delta^a_o
\end{equation}
where $r_o$ is a constant.

{\bf V. Boundary of the Photon Cloud}

The radiation zone described by the metric in (5) can be made continuous
with either the Schwarzschild metric describing the vacuum space-time outside
the radiation zone or with the Vaidya metric describing the zone of only 
directed radiations. In both cases we will express the exterior
also in null coordinates
\begin{equation}ds^2=-(1-{2M(u)\over r})du^2\pm 2dudr +r^2(d\theta^2+\sin^2{\theta}d\phi^2)
\end{equation}
For Schwarzschild space-time $M(u)=M_s$ where $M_s$ is a constant and
denotes Schwarzschild mass. If $\Sigma$ is the three surface boundary
separating the two regions of space-time than junction conditions have to
be satisfied across the boundary. The jump conditions$^7$  
basically require that
the metric $g_{ab}$ and the energy flux vector $T_{ab}n^b$ be continuous
across the boundary $\Sigma$ to which $n^a$ is the normal vector.

We would limit our consideration of jump conditions
only across the three surfaces $u=constant={\it T}$ and $r=constant =r_c$
keeping in mind that exterior metric is either Schwarzschild or Vaidya.
Below we consider matching conditions across the boundary $r=const.$
which is spacelike and the null boundary at $u=const.$

{\bf A. $\Sigma=r-r_c=0$}  

The junction conditions across the boundary $r=r_c=const.$, for the 
exterior metric tobe the either Vaidya metric consisting of only directed
radiations or vacuum Schwarzschild imply that both the metric and
the energy momentum flux be continuous across $r=r_c$. We have
\begin{equation}\Sigma=r-r_c=0\Rightarrow n_a=
{1\over \sqrt{g^{11}}}(0,1,0,0)\end{equation} 
Continuity of
metric for the form of metric in null coordinates given in (5) implies
\begin{equation}[A]_{r=r_c}= [C],\quad [\beta]_{r=r_c}=[C]\end{equation} 
where $[C]$ means continuity. 
The requirement that flux $T^{ab}n_b$ be continuous implies

\begin{equation}
{1\over \sqrt{g^{11}}}T^1_0=[C],\quad     
{1\over \sqrt{g^{11}}}T^1_1=[C]
\end{equation}

We now consider the cases when exterior metric is described by either
Vaidya or Schwarzschild solutions  

{\bf a. Exterior Vaidya metric}

The junction conditions across the boundary $r=r_c$ given ( 42) and
(43) above in case
of exterior metric being Vaidya
space-time  become
\begin{equation}[A]_{r=r_c}= 1-{2M(u)\over r_c},
\quad [\beta]_{r=r_c}=0\end{equation} 
\begin{equation}[(rA)']_{r=r_c}=1,\quad 
[\dot {(Ae^{-4\beta})}]_{r=r_c}=[{-2{dM\over dv}\over r_c}]\end{equation}
Since these conditions require the continuity of the derivatives of
functions with respect to $r$ and therefor it is not straight 
forward to obtain
matching conditions in general , hence we 
take the exact solution where $\beta$ given by equations (23) 
and exact solution by equations (24) to (26) with $n=const., r_o=const.$.
The junction condition
imply

\begin{equation} r_c=r_o\end{equation}
\begin{equation}M(u)+E(u) +{2\over a_o}=1-{2M(u)\over r_c}\end{equation}
\begin{equation}nM(u)+lE(u)+{8m\over a_o}=
{2M\over r_c}\end{equation}
where we have put $M(u)=(r_o)^nM_1(u)$ and $E(u)=(r_o)^lM_2(u)$.
The above equations relate the functions $M(u), E(u)$ to the Vaidya
mass $M(u)$.

{\bf b. Photon Gas Sphere in Vacuum}

We consider now the case of the static solution representing a photon cloud
and see whether the junction condition across the 
boundary $r=const.=r_c$ of the static photon cloud
can satisfied for a possible exterior Schwarzschild metric given in 
(38) with $M(u)=const.=M_s$.
The O'Brine and Synge junction condition given by (42) and (43) become
\begin{equation}[A]_{r=r_c}= 1-{2M_s\over r_c},
\quad [\beta]_{r=r_c}=0\end{equation} 
\begin{equation}[(rA)']_{r=r_c}=1 \end{equation}
Thus static solutions can be matched if the above are satisfied. As an 
illustration let us consider a simple exact solution
given in equations (23) to (26) with
$n=M_2(u)=0$ and $r_o=const., M_1(u)=const.=M_1$ we have $A$ and $\beta$ for
such a solution
$$e^{2\beta}= ({r\over r_o})^{{1\over 2}},\quad A=M_1 +{4r\over 7r_o}$$
The junction conditions given above are satisfied 
\begin{equation}r_c=r_o,\quad M_1=-{1\over 7},\quad r_o={14M_s\over 4}
\end{equation} 

Before closing the discussion on junction conditions we briefly mention
that all static solutions with only isotropic photon gas as given
by equations (36) to (38) do not satisfy the junction conditions
above as expected. Thus although a pure isotropic photon gas sphere can not
exist in a vacuum exterior, a mixture of directed radiation and
isotropic photon gas can be matched to the Schwarschild vacuum solution at
a sphere of radius $14M_s/4$. 
This means that a mixture
of both isotropic and directed radiations need not necessarily diffuse
through vacuum.

{\bf B. $\Sigma =u-{\it T=0}$, boundary of the radiation zone}

Let us consider  the boundary of the radiation zone at $u=T=constant$ and
exterior metric tobe Schwarzschild. 
The junction conditions then require that at the boundary $\Sigma=u-T=0$
with normal $n_a=(1,0,0,0)$ the metric
be continuous which means
\begin{equation}[A]_{u={\it T}}= (1-{2M_s\over r}),\quad [\beta]_{u={\it T}}=0
\end{equation} 
and the requirement that flux be continuous implies
\begin{equation}[T_{rr}]_{u={\it T}}=0\Rightarrow [\beta ']_{u={\it T}}=0,
\quad [T_{ur}]_{u={\it T}}=0
\Rightarrow [(rA)']_{u={\it T}}=1\end{equation}

Thus the junction conditions basically put restrictions on the choice
of the function $\beta$.
The metric in (5) is continuous with the
exterior Schwarzschild space-time across the boundary $u={\it T}$ provided
$[\beta]_{u={\it T}}=[\beta ']_{u={\it T}}=0$. 
This implies from equations (9) and  (17)
that at the boundary one of the solution of the homogeneous equation is
$[y=y_2(u,r)]_{u={\it T}}={1\over r}$ while the other 
$[y=y_1(u,r)]_{u={\it T}}={1\over r^2}$ 
and thus (18) gives
$A$ at the boundary  
\begin{equation}[A]_{u={\it T}}=1+{M_1({\it T})\over r^2}+
{M_2({\it T})\over r}\end{equation}
and therefore the boundary conditions imply
\begin{equation}M_2({\it T})=-2M_s,\quad M_1({\it T})=0\end{equation}
Thus all the solutions with function $\beta (u,r)$ such that 
$\beta (u={\it T},r)=\beta '(u={\it T},r)=0$ the junction conditions are 
satisfied. For specific examples of exact solution given in equations 
(30) to (32) these imply
that $\mu({\it T})=0$ and $M_2({\it T})=-2M_s,\quad M_1({\it T})=0$

{\bf APPENDIX A}

Because of the spherical symmetry one can decompose the spacetime metric
as a combination of two 2-subspaces i.e. we can write 
\begin{equation}ds^2=g_{ab}dx^{a}dx^{b}=g_{ij}dx^{i}dx^{j}+R^2
(d\theta^2+\sin^2{\theta}
d\phi^2)\end{equation}
\begin{equation}x^a=(x^i,\theta,\phi), a=0,1,2,3\quad i=0,1 \end{equation}
where $ x^i, i =0,1$ are arbitrary coordinates which span the
radial two space $(\theta, \phi)=const.$,
$g_{ij}(x^i)$ is the metric of the radial two space and $R=R(x^i)$ 
describes the area of the two spheres $x^i=const.$.

Spherical symmetry imposes restrictions on the energy tensor
$T^{ab}$. Thus there are two real spacelike eignvectors which
lie in the two space for which $dx^i=0$  
and the remaining in the two space for which $d\theta=d\phi=0$.
In the former case the two eignvalues are equal.
Thus the energy tensor $T^{ab}$ has only $T^{00}, T^{01}, T^{11}$, and
$T^{\theta}_{\theta}=T^{\phi}_{\phi}\rightarrow 
sin^2\theta T^{\theta\theta}=T^{\phi\phi}$ as non vanishing components.
Let 
$(p_{(\theta)}^a,p_{(\phi)}^a)$ be the two spacelike  
eignvectors of the stress energy tensor
which lie in the two space 
$x^i=const.$ with eignvalues $\lambda_{\theta}$
we have 
\begin{equation}p_{(\theta)}^a=
{1\over R}\delta^a_{\theta}, 
p_{(\phi)}^a={1\over R\sin{\theta}}
\delta^a_{\phi}\end{equation}
We further introduce a frame of two vectors which lie in the radial two space
namely a timelike vector $v^a$ and a null vector $k^a$ such that
\begin{equation}v^av_a=-1,\quad k^ak_a=k_ap_{(\theta)}^a=
k_ap_{(\phi)}^a=v_ap_{(\theta)}^a=
v_ap_{(\phi)}^a=0\quad \end{equation}
In general we can therefore express the stress energy tensor 
$T^{ab}$ as
\begin{equation}T^{ab}=A_ov^av^b+B_ok^ak^b+C_o(k^av^b+k^bv^a)+
\lambda_{\theta} p_{(\theta)}^a
p^b_{(\theta)} +\lambda_{\theta} 
p_{(\phi)}^ap^b_{(\phi)}
\end{equation}
where $(A_o,B_o,C_o)$ represent components of $T^{ab}$.
Since the null vector lies in the two radial space we write $k^a=Q^a+v^a$
where $Q^a$ is a spacelike unit vector orthogonal to $v^a$. By writing
\begin{equation}k^av^b+k^bv^a=2v^av^b+Q^av^b+Q^bv^a\end{equation}
and replacing $(k^av^b+k^bv^a)$ in  (60) we get $T^{ab}$ in the following
form
\begin{equation}T^{ab}=A_1v^av^b+B_1k^ak^b+C_1(Q^av^b+Q^bv^a)+\lambda_{\theta} g^{ab}
\end{equation}
Requiring that $T^a_a=0$ we get
\begin{equation}A_1=4\lambda_{\theta}\end{equation}
and we can therefore express in general
\begin{equation}T^{ab}={4\rho \over 3}v^av^b+{\rho\over 3}g^{ab}+
\sigma k^ak^b+q^av^b+q^bv^a\end{equation}
where we have changed the notations as 
$\lambda_{\theta}={\rho\over 3}, B=\sigma, q^a=C_1Q^a$ 
for the sake of physical
reasons only.

The general matter fields described in (60) for spherically symmetric
spacetime are physically realistic if they are either type I or II matter
fields$^4$. The type I fields are characterized by the existence of 
two distinct orthonormal real eignvectors in the radial two space while 
type II is characterized by the existence of 
one double null real eignvectors in the radial two space and can be
written as 
\begin{equation}T^{ab}=\lambda_0E_{(0)}^aE^b_{(0)}+\lambda_1E^a_{(1)}E^b_{(1)}
+\lambda_{\theta} p_{(\theta)}^ap^b_{(\theta)} +\lambda_{\theta} 
p_{(\phi)}^ap^b_{(\phi)}\end{equation}
\begin{equation}T_{ab}=-2\lambda_{II}{\it l}_{(a}m_{b)}+m_am_b
+\lambda_{\theta} p_{(\theta)}^ap^b_{(\theta)} +\lambda_{\theta} 
p_{(\phi)}^ap^b_{(\phi)}\end{equation}
where $E^a_{(0)}$ is the unit timelike eignvector othogonal to
unit spacelike  eignvector $E^a_{(1)}$ which lie in the radial two 
space with $\lambda_o$ and $\lambda_1$
as corresponding eignvalues in case of type I field in (65)  
and $({\it l}^a, m^a)$ are two real null vectors in the radial two space
in case of type II fields in (66).

All known observed physical fields are of type I except for the pure 
directed radiations  which are the only known physically realistic fields
belonging to type II fields. Therefore type I fields are of special interest.
Let us consider a type I spherically symmetric matter fields given in (65).
For matter cloud with a vanishing stress energy scalar $T=T^a_a=0$ 
imply
\begin{equation}\lambda_1-\lambda_0+2\lambda_{\theta}=0\Rightarrow 
\lambda_1=\lambda-\lambda
_{\theta},\quad \lambda_o=\lambda+\lambda_{\theta}\end{equation}

In case the eignvalues in the radial two space are
distinct that is $\lambda_o\ne\pm\lambda_1$ we
could use a set of a null vector $k^a$ and a timelike vector $v^a$ in the
radial two space such that
\begin{equation}E^a_{(1)}=\sqrt{{2\lambda_{\theta}\over \lambda}}v^a +
{1 \over 2}\sqrt{{2\lambda_{\theta}\over \lambda}}(1+{{2\lambda_
{\theta}\over \lambda}}){k^a\over k^av_a}\end{equation}
\begin{equation}E^a_{(o)}=-\sqrt{{2\lambda_{\theta}\over \lambda}}v^a +
{1 \over 2}\sqrt{{2\lambda_{\theta}\over \lambda}}(1-{{2\lambda_
{\theta}\over \lambda}}){k^a\over k^av_a}\end{equation}
The stress energy tensor for type I fields then can be expressed as
\begin{equation}T^{ab}={4\rho\over 3}v^av^b +{\rho\over 3}g^{ab} +\sigma k^ak^b\end{equation}
where we changed the notation $\lambda_{\theta}={\rho\over 3}$ 
for the sake of for physical reasons. 
In special cases where $\lambda_0=\pm\lambda_1$  
we can cast $T^{ab}$  in (65) as
\begin{equation}T^{ab}=\sigma (k^ak^b+q^{a}u^{b}+q^bu^a),\quad \hbox{for }
\lambda_0=\lambda_1\end{equation}
\begin{equation}T^{ab}={\rho\over 3}(4v^av^b +g^{ab}+ k^ak^b+
q^{a}u^{b}+q^bu^a),
\quad \hbox{for }
\lambda_0=-\lambda_1\end{equation}
where $k^a=q^a-v^a$ is a null vector, $v^a$ is a unit timelike vector 
which is orthogonal to spacelike unit vector $q^a$.

Though in this paper we would only deal with type I matter fields, however,
for the sake of completeness of
discussion we wish to mention that type II fields characterized by 
equation (66) can
be expressed in case $\lambda_{II}=0$ as
\begin{equation}T^{ab}=\sigma k^ak^b +
\lambda_{\theta} p_{(\theta)}^ap^b_{(\theta)} +\lambda_{\theta} 
p_{(\phi)}^ap^b_{(\phi)}\end{equation}
or in case $\lambda_{II}\ne 0$
\begin{equation}T^{ab}={4\rho \over 3}v^av^b+{\rho\over 3}g^{ab}+
(q^av^b+q^bv^a) 
\end{equation}

{\bf Appendix B.}

Here we give the necessary geometric quantities associated with the
metric
\begin{equation}ds^2=-Adu^2\pm 2e^{2\beta}dudr +r^2
(d\theta^2+\sin^2{\theta}d\phi^2)
,\quad u,r,\theta,\phi=0,1,2,3\end{equation}
The non vanishing metric components and 
Christoffel symbols associated with the metric
are 
\begin{equation}g^{11}=Ae^{-4\beta},\quad g^{01}=\pm e^{-2\beta},
\quad g^{22}=r^2\quad g^{33}=r^2\sin^2{\theta}
\end{equation}
\begin{equation}\Gamma^2_{12}=\Gamma^3_{13}={1\over r},\quad, 
\Gamma^2_{32}=\cot{\theta},
\Gamma^2_{33}=-\sin{\theta}\cos{\theta}\end{equation}
\begin{equation} \Gamma^1_{22}=\sin^{-2}{\theta}
\Gamma^1_{33}=-rg^{11}, \quad \Gamma^0_{22}=\sin^{-2}{\theta}
\Gamma^0_{33}=-rg^{01}\end{equation}
\begin{equation}\Gamma^0_{00}=2\dot \beta\pm {A'\over 2}e^{-2\beta}, 
\Gamma^1_{11}=2\beta ',
\quad \Gamma^1_{01}=\mp{A'e^{-2\beta}\over 2}\end{equation}
\begin{equation}\Gamma^1_{00}=\pm(2\dot \beta A-{\dot A\over 2})e^{-2\beta}+ 
{A'\over 2}Ae^{-4\beta}\end{equation}
The non vanishing components of the Ricci Tensor are
\begin{equation}R_{11}={4\beta '\over r}\end{equation}
\begin{equation}R_{22}=1-e^{-2\beta}(rAe^{-2\beta})'\end{equation}
\begin{equation}R_{00} 
=\pm e^{2\beta}\dot{({Ae^{-4\beta}\over r})}\pm Ae^{-2\beta}\dot\beta'
+{Ae^{-2\beta}\over 2}(A''e^{-2\beta}+2A'e^{-2\beta}({2\over r}-2\beta')
)\end{equation}
\begin{equation}R_{01} 
=-\dot\beta'
\mp{e^{-2\beta}\over 2}(A''+2A'({2\over r}-2\beta')
)\end{equation}

\begin{equation}R=\left (A''+{2A'\over r}(2-r\beta')+{2A\over r^2}(
1-4r\beta')\right )e^{-4\beta}
-{2\over r^2}\pm 4\dot\beta'\end{equation}

{\bf References}

\noindent
1. D. Kramer, H. Stephani, E. Herlt and M. MacCallum, 1980, Exact
Solutions of Einstein's Field Equations, Cambridge University Press.

\noindent
2. P. C. Vaidya Nature 171, 260, (1953).

\noindent
3. Dwivedi I H and Joshi P S 1989 Class Quantunm Grav 6 1599.;K. Lake
Phys. Rev. D 43 1416(1991)

\noindent
4. Hawking S W and Ellis G F R 1973 The Large Scale Structure of Space-Time
Cambridge University Press. p. 88.

\noindent

5. J. L. Synge,  1957 The Relativistic Gas (Amesterdam North Holland)
p. 37; Thorne K. S. Mon. Not. R. astr. Soc 194, 439 (1981);
Weinberg, S. Astrophys. J. 168, 175, (1972)

6. W. B. Bonor and P. C. Vaidya, Gen Relativ. Gravit. 1, 127, (1970);
B T Sullivan and W Israel Phys. Lett. 79A, 371, (1980).

7. Synge J. L. 1961 Relativity: The General Theory (Amesterdam: North Holland)
   P.39 

8. Perfect fluid solutions with $p=\rho/3$ has been discussed in cosmological 
context such as Friedmann models. See also Klein. O.
(1947) Ark. Mat. Astr. Pys. A 34 1. See 14.1.

\end{document}